\documentclass[preprint,aps,floats,showpacs]{revtex4}
\usepackage{graphicx}
\usepackage{epsfig}
\usepackage{pstricks}
\usepackage{pst-coil}
\usepackage{amsmath}

\def\be{\begin{equation}}
\def\ee{\end{equation}}
\def\bea{\begin{eqnarray}}
\def\eea{\end{eqnarray}}
\begin{document}
\begin{flushright}
hep-ph/0501069, v2
\end{flushright}
\title{\bf Heavy neutrino mass hierarchy from leptogenesis 
in left-right symmetric models with spontaneous CP-violation}

\author{Narendra Sahu}
\email{narendra@phy.iitb.ac.in}
\author{S.~ Uma Sankar}
\email{uma@phy.iitb.ac.in}
\affiliation{Department of Physics, Indian Institute of 
Technology, Bombay, Mumbai 400076, India}                              
\today

\begin{abstract}
We consider left-right symmetric model with spontaneous  
CP-violation. The Lagrangian of this model is CP invariant
and the Yukawa couplings are real. Due to spontaneous breaking 
of the gauge symmetry, some of the neutral Higgses acquire 
complex vacuum expectation values, which lead to CP-violation.
In the model considered here, we identify the neutrino Dirac 
mass matrix with that of Fritzsch type charged lepton mass 
matrix. We assume a hierarchical spectrum of the right handed 
neutrino masses and derive a bound on this hierarchy by assuming 
that the decay of the lightest right handed neutrino produces the 
baryon asymmetry via the leptogenesis route. It is shown 
that the mass hierarchy we obtain is compatible with the 
current neutrino oscillation data.  
\end{abstract}
\maketitle
%%%%%%%%%%%%%%%%%%%%%%%%%%%%%%%% 
\section{Introduction}
Present low energy neutrino oscillation data~\cite{solar,atmos,
kamland} are elegantly explained by neutrino oscillation hypothesis
with very small masses ($\leq 1$ eV) for the light neutrinos. 
These masses can be either Dirac or Majorana. See-saw mechanism
is an elegant technique to generate small Majorana masses for
light neutrinos without fine tuning~\cite{ge-ra-sl-ya}. This can 
be achieved by introducing right handed neutrinos into the electroweak
model, which are invariant under all gauge transformations. The
Majorana masses of these right handed neutrinos are free parameters
of the model and are expected to be either at TeV scale~\cite{
sahu&yajnik_prd.04} or at a higher scale~\cite{davidson&ibarra.02,
buch-bari-plum.02,antusch.04,sahu&uma_prd.04}.

In the simplest scenario a right handed neutrino per 
generation is added. They are coupled to left handed neutrinos 
via Dirac mass matrix ($m_{D}$) which is assumed to be similar 
to charged lepton mass matrix. The consequent type-I seesaw 
mechanism~\cite{ge-ra-sl-ya} gives rise to Majorana mass matrix of 
the light neutrinos of the form 
\be
m_\nu=-m_DM_R^{-1}m_D^T.
\label{type-I}
\ee
The goal of the present neutrino oscillation experiments 
is to determine the nine degrees of freedom of the above 
equation (\ref{type-I}). These are given by three light 
neutrino masses, three mixing angles and three phases 
which include one Dirac and two Majorana. At present the 
neutrino oscillation experiments are able to measure the two mass 
square differences, the solar and the atmospheric, and three 
mixing angles with varying degrees of precision, while there is 
no information about the phases. Moreover, it is
difficult to constrain the parameters of the right handed neutrinos
from low energy neutrino data. However, an early attempt
~\cite{smirnov_jhep} was made by inverting the seesaw formula 
(\ref{type-I}). 

Baryogenesis via leptogenesis~\cite{fukugita.86} provides an 
attractive scenario to link the physics of right handed neutrino 
sector with the low energy neutrino data. In this scenario, a 
lepton (L) asymmetry is produced first which is then 
transformed to baryon (B) asymmetry of the Universe via the high 
temperature behavior of the $B+L$ anomaly of the Standard Model 
($SM$)~\cite{krs.86}. Most proposals along these lines rely on 
the out of equilibrium decay of heavy Majorana neutrinos to 
generate the L-asymmetry~\cite{fukugita.86,luty.92,plumacher.96}. 
In these modifications of SM, the $B-L$ conservation is ad hoc. 

An alternative is to consider leptogenesis~\cite{mohapatra_prd.92,
tyatgat_plb.93} within left-right symmetric model~\cite{
leftright_group} where $U(1)_{B-L}$ is a gauge symmetry. Because 
$B-L$ is a gauge charge of the model, no primordial $B-L$ can 
exist. Further, the rapid violation of $B+L$ 
conservation by the anomaly due to high temperature sphaleron 
fields erases any $B+L$ generated earlier. Thus the lepton asymmetry 
must be produced entirely during or after the $B-L$ symmetry 
breaking phase transition. The Higgs sector of this model is very 
rich which consists of two triplets $\Delta_L$ and $\Delta_R$ and 
a bi-doublet $\Phi$. In contrast to type-I models, in the present 
case the vacuum expectation value (VEV) of the triplet 
$\Delta_L$ provides an additional mass, $m_L$, to the light 
Majorana neutrino mass matrix~\cite{magg-wet.80,wett.81,
moh-senj.81,laz-shf-wet.81,moha-susy-book.92}
\be
m_{\nu}=m_L-m_D M_R^{-1} m_D^T=m_\nu^{II}+m_\nu^{I},
\label{type-II}
\ee
where the two terms on the right hand side of above equation 
are called type-II and type-I respectively. 

In the present work, we consider a left-right symmetric model 
in which CP-violation occurs via spontaneous symmetry 
breaking~\cite{senjanovic.79,mas_npb,bas_npb,deshpande.91}. 
The Lagrangian of the model is CP invariant which demands that 
all the Yukawa couplings should be real. CP violation occurs
via the complex vacuum expectation values (VEVs) of the neutral 
Higgses in the model. In the present case, there are four complex 
neutral scalars, all of which can acquire complex VEVs. 
However, the global $U(1)$ symmetries associated 
with $SU(2)_L$ and $SU(2)_R$ gauge groups allow two of the 
phases to be set to zero. Using the remnant U(1) symmetry
related to $SU(2)_R$, one phase choice is made to
make the VEV of $\Delta_R$, and hence the mass matrix of right 
handed neutrinos, real. The phase associated with the other
$U(1)$ symmetry can be chosen to achieve two different types
of simplification of neutrino mass matrix. In the 
{\it type-II choice}, the $m_\nu^I$ is made real leaving 
the CP-violating phase purely with $m_\nu^{II}$. In this phase 
convention, we derive a lower bound on the mass scale of $N_1$ 
from the leptogenesis constraint by assuming a normal mass hierarchy 
in the right handed neutrino sector. In the {\it type-I phase 
choice}, only the type-I term contains CP-violating phase 
leaving type-II term real. This allows us to derive an upper bound on 
the heavy neutrino mass hierarchy from the leptogenesis constraint. 

The early analyses~\cite{mas_npb,bas_npb,deshpande.91} show that,  
the vacuum alignment of the most general Higgs potential 
in the left-right symmetric model requires both the phases to 
be very tiny ${\cal O}(m_W/v_R)$ ($v_R = \langle\Delta_R\rangle$) 
and hence there is no observable CP-violation. However, this analyses 
assumed that the VEVs of the two neutral scalars in the 
bidoublet $\Phi$ are of the same order of magnitude. On the 
other hand, requiring one of the bidoublet VEVs to be much 
smaller than the other~\cite{senjanovic.79} allows the phase
associated with the triplet VEV to be large~\cite{footnote1}.
Later analyses worked out scenarios where the above 
conclusion~\cite{footnote1} was explicitly demonstrated~
\cite{bar_rod}. These papers also showed that the choice 
$v_R \geq 10^8$ GeV, suppresses flavour changing neutral 
currents adequately. This is in accord with our result, shown 
in section V, that the present $B$-asymmetry of the Universe 
also requires a similar magnitude of $v_R$.
     
Rest of the paper is organised as follows. In section II, 
we briefly recapitulate the left-right symmetric model with 
spontaneous CP-violation. In section III, we derive an upper 
bound on the CP-asymmetry in left-right symmetric models by 
keeping both type-I and type-II terms in the mass matrix of light 
neutrinos. We identify the neutrino Dirac mass matrix with that of 
charged lepton mass matrix~\cite{ge-ra-sl-ya}. Further we choose 
this matrix to be of Fritzsch type~\cite{fritzsch.79}. With these 
assumptions, in section IV, we show that a successful lepton 
asymmetry can be created for a reasonable mass hierarchy among 
right handed neutrinos and derive bounds on this hierarchy. In 
section V, we demonstrate that this hierarchy is compatible with 
the current neutrino oscillation data. Section VI contains our 
summary and conclusions.

%%%%%%%%%%%%%%%%%%%%%%%%%%%%%%%%%%%%%%%%%%%%%%%%%%%%
\section{Left-right symmetric model and spontaneous 
CP-violation}
%%%%%%%%%%%%%%%%%%%%%%%%%%%%%%%%%%%%%%%%%%%%%%%%%%%%
In the left-right symmetric model the right handed 
charged lepton of each family, which was a singlet under the 
$SM$ gauge group $SU(2)_L \otimes U(1)_Y$, gets a new partner 
$\nu_R$. These two form a doublet under the $SU(2)_R$ of the 
left-right symmetric gauge group $SU(2)_L \otimes SU(2)_R \otimes 
U(1)_{B-L}$. Similarly, in the quark sector, the right handed up and 
down quarks of each family, which were singlets under $SM$ gauge 
group, combine to form a doublet under $SU(2)_R$. 

The Higgs sector of the model consists of two triplets 
$\Delta_L$ and $\Delta_R$ and a bidoublet $\Phi$, which contains
two copies of $SM$ Higgs. Under $SU(2)_L\otimes SU(2)_R 
\otimes U(1)_{B-L}$ the field content and the quantum numbers 
of the Higgs fields are given as 
\begin{eqnarray}
\Phi &=&\begin{pmatrix}
\phi_{1}^{0} & \phi_{1}^{+}\\
\phi_{2}^{-} & \phi_{2}^{0}
\end{pmatrix} \sim (1/2,1/2,0)\\
\Delta_{L} &=& \begin{pmatrix}
\delta_{L}^{+}/\sqrt{2} & \delta_{L}^{++}\\
\delta_{L}^{0} & -\delta_{L}^{+}/\sqrt{2}
\end{pmatrix}\sim (1,0,2)\\
\Delta_{R} &=& \begin{pmatrix}
\delta_{R}^{+}/\sqrt{2} & \delta_{R}^{++}\\
\delta_{R}^{0} & -\delta_{R}^{+}/\sqrt{2}
\end{pmatrix}\sim (0,1,2).
\end{eqnarray} 

To achieve the correct phenomenology, the various Higgs multiplets in 
the model should have the following VEVs,
\be
\langle \Delta_{R}\rangle =\begin{pmatrix}
0 & 0\\
v_R e^{i\theta_R} & 0 \end{pmatrix},
\label{right_vev}
\ee
\be
\langle \Phi \rangle =\begin{pmatrix}
k_1 e^{i\alpha} & 0\\
0 & k_2 e^{i\beta}\end{pmatrix},
\label{dirac_vev}
\ee
and
\be
\langle \Delta_{L}\rangle =\begin{pmatrix}
0 & 0\\
v_L e^{i\theta_L} & 0 \end{pmatrix}.
\label{left_vev}
\ee
The electric charge of the fields is given by  
\be 
Q=T^3_L + T^3_R+ \frac{1}{2}(B-L).
\ee 
In the above $v_L$, $v_R$, $k_1$ and $k_2$ are real parameters 
and the electroweak symmetry breaking scale $v=174$ GeV is given by 
$v^2=k_1^2+k_2^2$. Further we require that $v_L\ll v\ll v_R$. The 
requirement of the spontaneous breakdown of parity gives rise to 
\be
v_L v_R=\gamma (k_1^2+k_2^2)=\gamma v^2,
\ee 
where $\gamma$ is parameter which is a function of the quartic couplings 
in the Higgs potential. 

The minimisation of the most general Higgs potential involving 
$\Delta_L, \Delta_R$ and $\Phi$ was studied in refs. \cite{mas_npb, 
bar_rod}. The relations between the various couplings, for which 
the above set of VEVs are generated, were derived. In this 
scenario, the gauge symmetry $SU(2)_L\times SU(2)_R \times U(1)_{B-L}$ 
is broken to $U(1)_{em}$ in a single step. Thus the $CP$-violating 
phases come into existence at the same scale where the left-right 
symmetry is broken. Since $v\ll v_R$, the $SM$ symmetry is present as 
an approximate symmetry at the scale where symmetry breaking occurs. 

An attractive alternative to the single step symmetry breaking is 
the phenomenon of inverse symmetry breaking \cite{weinberg_prd.74,
moha_sen_prl.79}. This phenomenon usually occurs in models with 
multiple HIggs representations. The zero temperature Higgs potential
of the model can be chosen so that all the neutral Higgses acquire
non-zero VEVs. To consider the pattern of symmetry breaking at high
temperatures, a temperature dependent correction is added to the 
Higgs potential and the potential is minimized. At high temperatures,
it was shown that some of the VEVs grow with the temperature
\cite{moha_sen_prl.79,moh_sen_prd.79}. In ref. \cite{moh_sen_prd.79},
this mechanism was demonstrated for a left-right symmetric model
with two Higgs bidoublets. The relevance of this mechanism to the 
model considered here is being studied and will be reported soon.

The fermions get their masses via Yukawa couplings. The 
Lagrangian for one generation of quarks and leptons is 
\bea
-\mathcal{L}_{yuk} &=& \tilde{h}_q\bar{q}_L\Phi q_R+
\tilde{g}_q\bar{q}_L\tilde{\Phi}q_R+ \tilde{h}_l\bar{\ell}_L\Phi l_R+
\tilde{g}_l\bar{\ell}_L\tilde{\Phi}l_R \nonumber \\
& &+if (\ell_L^T C\tau_2\Delta_L \ell_L
+\ell_R^T C\tau_{2}\Delta_{R}\ell_R) + H.c.\,
\label{yukawa}
\eea 
where q and $\ell$ are quark and lepton doublets,
$\tilde{\Phi}=\tau_2 \Phi^*\tau_2$ and C is the Dirac charge
conjugation matrix. Further the Majorana Yukawa coupling
$f$ is the same for both left and right handed neutrinos to
maintain the discrete $L\leftrightarrow R$ symmetry.

Substituting the complex VEVs (\ref{right_vev}), (\ref{dirac_vev}) 
and (\ref{left_vev}) in (\ref{yukawa}) we obtain fermion masses to 
be
\bea
M_f &=& (\tilde{h}_q k_1 e^{i\alpha}+
\tilde{g}_q k_2 e^{i\beta})\bar{u}_L u_R+(\tilde{h}_q k_2 
e^{i\beta}+\tilde{g}_q k_1e^{i\alpha})\bar{d}_L d_R \nonumber\\
& &+(\tilde{h}_l k_1e^{i\alpha}+\tilde{g}_l k_2 e^{i\beta})
\bar{\nu}_L \nu_R+(\tilde{h}_l k_2 e^{i\beta}+\tilde{g}_l 
k_1e^{i\alpha})\bar{e}_L e_R \nonumber\\
& &+ f(\nu_L^T C v_L e^{i\theta_L}\nu_L +
\nu_R^T C v_R e^{i\theta_R}\nu_R)+H.C.
\label{eff-yukawa}
\eea     
Generalising the above equation (\ref{eff-yukawa}) for three 
generation of matter fields we get the up and down quark mass 
matrices to be
\be
(M_u)_{ij}=(\tilde{h}_q)_{ij} k_1 e^{i\alpha}+(\tilde{g}_q)_{ij}
k_2 e^{i\beta}~~\mathrm{and}~~
(M_d)_{ij}=(\tilde{h}_q)_{ij} k_2 e^{i\beta}+(\tilde{g}_q)_{ij} 
k_1 e^{i\alpha}.
\label{quark-mass}
\ee
We assume~\cite{deshpande.91,ball.00} $k_1/k_2\sim m_t/m_b$.
In the see-saw mechanism, the Dirac mass matrix of the neutrinos
is assumed to be similar to the mass matrix of the charged leptons.
For $k_2 \ll k_1$, and assuming $\tilde{h}_l\sim \tilde{g}_l$ in 
equation (\ref{eff-yukawa}), the Dirac mass matrix of the neutrinos 
to a good approximation becomes $\tilde{h}_l k_1 e^{i\alpha}$.
Thus neglecting $k_2$ terms, the masses of three generations of
neutrinos are given by
\be
(M_{\nu})_{ij}=\bar{\nu}_{L_i}k_1 e^{i\alpha}(\tilde{h}_l)_{ij}\nu_{R_j}
+f_{ij}(v_L e^{i\theta_L} \nu_{L_i}^T C \nu_{L_j}
+v_R e^{i\theta_R}\nu_{R_i}^T C\nu_{R_j}) + H.C.
\ee
The Majorana Yukawa coupling matrix $f_{ij}$ is real and symmetric
and hence can be diagonalized by an orthogonal transformation on
$\nu_R$
\be
N_R=O_R^T \nu_R.
\ee
In this basis, we have
\bea
O_R^T f O_R &=& f_{dia}\label{ortho_trans_f},\\  
          h &=& \tilde{h} O_R\label{ortho_trans_h}.
\eea
In the transformed basis we get the mass matrix for the neutrinos
to be
\be
\begin{pmatrix}
f v_L e^{i\theta_L} & k_1 e^{i\alpha}h\\
k_1e^{i\alpha} h^T & f_{dia}v_Re^{i\theta_R} \end{pmatrix}.
\ee
Diagonalising the neutrino mass matrix into $3\times 3$ blocks
we get the light neutrino mass matrix to be
\be
m_{\nu} = f v_{L} e^{i\theta_L} - \frac{k_1^2}{v_R}
(h f_{dia}^{-1} h^T)e^{i(2\alpha-\theta_R)}
\label{see-saw}
\ee   

Notice that the Lagrangian (\ref{yukawa}) is invariant under the 
following unitary transformations of the fermion and Higgs fields, 
\bea
\psi_L\longrightarrow U_L\psi_L~~ & \mathrm{and}~~ & 
\psi_R\longrightarrow U_R\psi_R, \\
\Phi\longrightarrow U_L\Phi U_R^{\dagger}~~ & \mathrm{and}~~ &  
\tilde{\Phi}\longrightarrow U_L\tilde{\Phi} U_R^{\dagger}\\
\Delta_L \longrightarrow U_L\Delta_L U_L^{\dagger}~~ 
& \mathrm{and}~~ & \Delta_R \longrightarrow U_R\Delta_R 
U_R^{\dagger},  
\eea
where $\psi_{L,R}$ is a doublet of quark or lepton fields. 
The invariance under $U_L$ is the result of the remnant 
global $U(1)$ symmetry which remains after the breaking of
the gauge symmetry $SU(2)_L$ and similarly for $U_R$.
The matrices $U_L$ and $U_R$ can be parametrized as 
\be
U_L=\begin{pmatrix}
e^{i\gamma_L}&0\\
0& e^{-i\gamma_L}\end{pmatrix}~~ \mathrm{and}~~ 
U_R=\begin{pmatrix}
e^{i\gamma_R}&0\\
0& e^{-i\gamma_R}\end{pmatrix}.
\ee
By redefining the phases of the fermion fields we can rotate
away two of the phase degrees of freedom from the scalar
sector of the theory. Thus only two of the four phases of
Higgs VEVs have phenomenological consequences. 
Under these unitary transformations, the VEVs (\ref{right_vev}), 
(\ref{dirac_vev}) and (\ref{left_vev}) become 
\be
\langle \Delta_{R}\rangle =\begin{pmatrix}
0 & 0\\
v_R e^{i(\theta_R-2\gamma_R)} & 0 \end{pmatrix},
\label{right_r_vev}
\ee
\be
\langle \Phi \rangle =\begin{pmatrix}
k_1 e^{i(\alpha+\gamma_L-\gamma_R)} & 0\\
0 & k_2 e^{i(\beta-\gamma_L+\gamma_R)}\end{pmatrix},
\label{dirac_r_vev}
\ee 
and 
\be
\langle \Delta_{L}\rangle =\begin{pmatrix}
0 & 0\\
v_L e^{i(\theta_L-2\gamma_L)} & 0 \end{pmatrix}.
\label{left_r_vev}
\ee
We choose $\gamma_R = \theta_R/2$ so that the masses of 
the right handed neutrinos are real. The light neutrino mass 
matrix (\ref{see-saw}) then becomes 
\bea
m_{\nu} &=& f v_{L} e^{i(\theta_L-2\gamma_L)}-\frac{k_1^2}{v_R}
(h f_{dia}^{-1} h^T)e^{i(2\alpha+2\gamma_L-\theta_R)}
\label{rot-seesaw}\\
&=& m_{\nu}^{II}+m_{\nu}^{I}
\eea

Conventionally, in equation (\ref{rot-seesaw}), $\gamma_L$ was 
chosen to be $-\alpha + \theta_R/2$ ~\cite{deshpande.91,
mahanthapa.04}. This makes $m_\nu^I$ real leaving the imaginary
part purely in $m_\nu^{II}$.
We call this {\it type-II phase} choice. The 
light neutrino mass matrix, with this phase choice, is 
\be
m_\nu=f v_{L} e^{i\theta'_L}-\frac{k_1^2}{v_R}(h f_{dia}^{-1} h^T),
\label{con_seesaw}
\ee
where $\theta'_L=(\theta_L-\theta_R+2\alpha)$. On the other hand, 
by choosing $\gamma_L = \theta_L/2$ in equation (\ref{rot-seesaw}) 
$m_\nu^{II}$ can be made real, with the phase occuring purely in 
$m_\nu^I$. We call this {\it type-I phase} choice. Consequently 
the light neutrino mass matrix (\ref{rot-seesaw}) becomes 
\be
m_\nu = f v_{L}-\frac{k_1^2}{v_R}e^{i\theta'_R}
(h f_{dia}^{-1} h^T)\,
\label{chosen_seesaw}
\ee
where $\theta'_R=(\theta_L-\theta_R+2\alpha)$. Note that the 
authors in ref.~\cite{mahanthapa.04} displayed the possibility 
of leptogenesis through the type-II choice only. With this choice,
they related the magnitude of CP violation in leptogenesis to the
magnitude of CP violation possible in neutrino oscillations in 
certain models. However, in the present work we consider two 
distinct phase choices, i.e. type-I and type-II, and consider
the implication of each choice to leptogenesis. The CP-violating parameter 
$\epsilon_1$ which gives rise to the lepton asymmetry is independent 
of the phase choice. However, the theoretical 
upper bound on $\epsilon_1$ is not a physical parameter of the 
theory and can depend on the choice of phases as we see in 
the next section. In numerical calculations, we take into account 
the consistency of the bounds coming from the different phase 
choices.

Diagonalization of the light neutrino mass matrix $m_{\nu}$,
through lepton flavour mixing matrix $U_{PMNS}$~\cite{mns-matrix},
gives us three light Majorana neutrinos. Its eigenvalues are
\be
U_{L}^T m_{\nu} U_{L}=dia(m_{1}, m_{2}, m_{3}),
\label{diag}
\ee
where $m_1$, $m_2$ and $m_3$ are the absolute masses of light
Majorana neutrinos and are chosen to be real.

\section{Upper bound on CP-asymmetry in left-right 
symmetric models}
%%%%%%%%%%%%%%%%%%%%%%%%%%%%%%%%%%%%%%%%%%%%%%%%%%%%%%% 
We assume that the lepton asymmetry of the Universe is 
produced by the CP-violating decay of the heavy right 
handed Majorana neutrinos to standard model leptons ($l$) 
and Higgs ($\phi$). We also assume a normal mass hierarchy for 
the heavy Majorana neutrinos. In this scenario while the 
heavier neutrinos, $N_2$ and $N_3$, decay, the lightest of 
heavy Majorana neutrinos is still in thermal equilibrium. 
Any asymmetry, thus, produced by the decay of $N_2$ and 
$N_3$ will be erased by the lepton number violating interactions 
mediated by $N_1$. Therefore, the final lepton asymmetry 
is given only by the CP-violating decay of $N_1$. The 
CP-asymmetry, thus, is given by
\be
\epsilon_1=\epsilon_1^I+\epsilon_1^{II},
\label{epsilon}
\ee                                 
where the contribution to $\epsilon_1^I$ comes from the
interference of tree level, self-energy correction and the
one loop radiative correction diagrams involving the heavier
Majorana neutrinos $N_2$ and $N_3$. This contribution is same 
as in type-I models~\cite{davidson&ibarra.02,
buch-bari-plum.02} and is given by
\be
\epsilon_1^{I}=\frac{3M_1}{16\pi v^2}\frac{\sum_{i,j}Im
\left[ h^T_{1i}(m_{\nu}^I)_{ij}h_{j1}\right]}
{(h^T h)_{11}}.
\label{epsilonI}
\ee
On the other hand the contribution to $\epsilon_1^{II}$ in
equation (\ref{epsilon}) comes from the interference of tree
level diagram and the one loop radiative correction
diagram involving the virtual triplet $\Delta_L$. It is given
by~\cite{hamb-senj.03,antusch.04}
\be
\epsilon_1^{II}=\frac{3M_1}{16\pi v^2}\frac{\sum_{i,j}Im
\left[ h^T_{1i}(m_{\nu}^{II})_{ij}h_{j1}\right]}
{(h^T h)_{11}}. 
\label{epsilonII}
\ee 
The total CP-asymmetry is therefore given by 
\be
\epsilon_1=\frac{3M_1}{16\pi v^2}\frac{\sum_{i,j}Im
\left[ h^T_{1i}(m_{\nu}^I+m_{\nu}^{II})_{ij}h_{j1}\right]}
{(h^T h)_{11}}.
\label{total-epsilon}
\ee 
From equation (\ref{total-epsilon}), we see that the physical 
observable $\epsilon_1$ is not affected by the choice of phases. 
In the following, we use bound on $\epsilon_1$ from the observed 
baryon asymmetry to obtain bounds on right-handed neutrino 
masses for the two different phase choices.

\subsection{The type-II choice of phases}
%%%%%%%%%%%%%%%%%%%%%%%%%%%%%%%%%%%%%%%%%%%%%%%%%%%%%%%%%%% 
In this choice of phases the type-I mass term is 
real. The only source of CP-violation in the light neutrino mass 
matrix $m_\nu$ lies in the type-II mass term. Thus in this 
case $\epsilon_1^{I}=0$ because of both h and $m_\nu^I$ are real. 
The total CP-asymmetry in this choice of phases is therefore 
given by 
\bea
\epsilon_1 &=& \epsilon_1^{II}\nonumber\\
&=& \frac{3M_1v_L}{16\pi v^2} \frac{(h^T f h)_{11}}
{(h^T h)_{11}}Im (e^{i\theta'_L}).
\label{choice-2-cp}
\eea  
Using (\ref{ortho_trans_f}) and (\ref{ortho_trans_h}) in 
equation (\ref{choice-2-cp}) we get 
\be
\epsilon_1=\frac{3M_1v_L}{16\pi v^2}\frac{ 
\sum_if_i(O_R^Th)_{i1}^2}{\sum_i(O_R^Th)_{i1}^2}sin\theta'_L,
\label{choice-2-cpasym}
\ee
where $f_i=(M_i/v_R)$. Up to a first order approximation it 
is reasonable to assume that $\sum_if_i\approx 1$. In this 
approximation the theoretical upper bound on the CP-asymmetry 
(\ref{choice-2-cpasym}) is given by~\cite{davidson&ibarra.02,
buch-bari-plum.02,antusch.04,sahu&uma_prd.04}  
\be
\epsilon_{1,max}=\frac{3M_1v_L}{16\pi v^2}.
\label{choice-2-max}
\ee
Thus, for type-II phase choice, a bound on $\epsilon_1$ 
leads to a bound on $M_1$.
\subsection{The type-I choice of phases}
%%%%%%%%%%%%%%%%%%%%%%%%%%%%%%%%%%%%%%%%%%%%%%%%%%%%%%%%%%%
In the type-I choice of phases the type-II mass term is real. 
Hence the CP-violation comes through the type-I mass term only. 
The total CP-asymmetry in this case is therefore given by
\bea 
\epsilon_1 &=& \epsilon_1^I\nonumber\\
&=& \frac{3M_{1}k_1^2}{16\pi v^2 v_R} \frac{(h^Th f_{dia}^{-1}
h^Th)_{11}}{(h^T h)_{11}}Im (e^{-i\theta_R'}).
\label{cpasym2}
\eea

Let us consider the type-I term of the light neutrino mass matrix
\bea
m_\nu^I &=& m_{\nu}-m_{\nu}^{II}\nonumber\\
      &=& -\frac{k_1^2}{v_Rx'} h f_{dia}^{-1} h^T e^{-i \theta_R'}.
\label{fakemass}
\eea
We can find a diagonalising matrix 
$U={\cal{O}}U_{phase}$ for $m_\nu^I$ such that 
\be
U^T m_\nu^I U \equiv -D_{m_I}=-dia(m_{I_{1}}, m_{I_{2}}, m_{I_{3}})
\label{fakedia}
\ee
where $(m_{I_{1}}, m_{I_{2}}, m_{I_{3}})$ are real by choosing
$U_{phase}=e^{i\theta_R'/2}$. Therefore, from equation
(\ref{fakedia}) we have 
\be
D_{m_{I}}=\frac{k_1^2}{v_R}{\cal{O}}^T \left(h f_{dia}^{-1}
h^T\right){\cal{O}}.
\label{mfdia}
\ee 
Using (\ref{mfdia}) in equation (\ref{cpasym2}) the CP-asymmetry
$\epsilon_1$ can be rewritten as
\bea
\epsilon_{1} &=& \frac{3M_{1}}{16\pi v^{2}}\frac{
\sum_{i}\left[(h^T{\cal{O}})_{1i} D_{m_{I_{ii}}} ({\cal{O}}^Th)_{i1}
\right]}{\sum_i \left[(h^T{\cal{O}})_{1i}({\cal{O}}^T h)_{i1}\right]}
Im (e^{-i\theta_R'})\nonumber\\
&=& \frac{3M_{1}}{16\pi v^{2}}\frac{
\sum_{i}m_{I_i}({\cal{O}}^Th)_{i1}^2}{\sum_i({\cal{O}}^Th)_{i1}^2)}
Im (e^{-i\theta_R'}).
\label{cpasym3}
\eea 
In the above equation (\ref{cpasym3}) the theoretical upper bound on 
CP-asymmetry is thus given by~\cite{davidson&ibarra.02,
buch-bari-plum.02}
\be
|\epsilon_{1, max}|=\frac{3M_{1}}{16\pi v^{2}}\sum_i m_{I_i}.
\label{cpasym-max}
\ee
In the equation (\ref{cpasym-max}) $m_{I}$s are the eigenvalues
of the matrix $m_\nu^I$ and {\it are not the physical light 
neutrino masses}. As we saw above, the relation between neutrino
masses and the theoretical upper bound on $\epsilon_1$ is phase
choice dependent.

It is desirable to express the $\epsilon_{1, max}$ 
in terms of physical parameters. In order to calculate the 
$m_{I}$s we will assume a hierarchical texture of Majorana 
coupling 
\be
f_{dia}= \frac{M_{1}}{v_{R}}\begin{pmatrix}1 & 0 & 0\\
0 & \alpha_{A} & 0\\
0 & 0 & \alpha_{B} \end{pmatrix},
\label{fdia-texture}
\ee
where $1 \ll \alpha_{A}=(M_{2}/M_{1}) \ll \alpha_{B}=
(M_{3}/M_{1})$. We identify the neutrino Dirac Yukawa 
coupling $h$ with that of charged leptons~\cite{ge-ra-sl-ya}. 
We assume $h$ to be of Fritzsch type~\cite{fritzsch.79}  
\be
h=\frac{(m_{\tau}/v)}{1.054618}
\begin{pmatrix}
0 & a & 0\\
a & 0 & b\\
0 & b & c\end{pmatrix}.
\label{h-texture}
\ee
We make this assumption because Fritzsch mass matrices are 
well motivated phenomenologically. By choosing the values 
of $a,b$ and $c$ suitably one can get 
the hierarchy for charged leptons and quarks. In particular 
~\cite{fritzsch.79} 
\begin{equation}
a=0.004,~~~ b=0.24 ~~~{\rm and}~~~ c=1 
\label{abcvalues}
\end{equation}
can give the 
mass hierarchy of charged leptons. For this set of values 
the mass matrix $h$ is normalized with respect to the 
$\tau$-lepton mass. The set of values of $a, b$ and $c$ 
are roughly in geometric progression. They can be expressed  
in terms of the electro-weak gauge coupling $\alpha_{w}= 
\frac{g^{2}}{4\pi}=\frac{\alpha}{sin^{2}\theta_{w}}$. In 
particular $a=2.9 \alpha_{w}^{2}$, $b=6.5 \alpha_{w}$ 
and $c=1$. Here onwards we will use these set of values 
for the parameters of $h$. Using equation (\ref{fdia-texture}) 
and (\ref{h-texture}) in equation (\ref{mfdia}), we now get 
\bea
D_{mI} &=& \frac{v^{2}}{v_{R}}
\left(h f_{dia}^{-1}h^T\right)_{dia}\nonumber\\
 &\simeq & \frac{v^{2}}{M_{1}} \frac{(m_{\tau}/v)^{2}}
{(1.054618)^{2}}
\begin{pmatrix}
0& 0 & 0\\
0 & A & 0\\
0 & 0 & B\end{pmatrix},
\label{eigenvalue}
\eea   
where the eigenvalues A and B are functions of $\alpha_{A}$ 
and $\alpha_{B}$ and their sum is given by 
\be
A+B = \frac{1}{2}\left[a^{2}+\frac{1}{\alpha_{A}}(a^{2}+b^{2}) 
+\frac{1}{\alpha_{B}}(b^{2}+c^{2})\right].
\label{eigenvalue-sum}
\ee
Using equation (\ref{eigenvalue}) we can write 
the maximum value of CP-asymmetry (\ref{cpasym-max})  
\bea
\epsilon_{1, max} &=& \frac{3 M_{1}}{16\pi v^{2}}
(m_{I2}+m_{I3})\nonumber\\
&=& \frac{3}{16\pi}\frac{(m_{\tau}/v)^{2}}
{(1.054618)^{2}}(A+B).
\label{cpasym-max1}
\eea
Thus we see that, in type-I choice of phases, the leptogenesis 
parameter $\epsilon_1$ constrains the hierarchy parameters 
$\alpha_A$ and $\alpha_B$. In the following two sections, we
will obtain numerical bounds on $\alpha_A$ and $\alpha_B$ 
in a manner consistent with the bound $M_1$ coming from the
type-II phase choice.

%%%%%%%%%%%%%%%%%%%%%%%%%%%%%%%%%%
\section{Estimation of Lepton Asymmetry}
%%%%%%%%%%%%%%%%%%%%%%%%%%%%%%%%%%%%%%%%%%
A net $B-L$ asymmetry is generated when the gauge symmetry 
is broken. A partial $B-L$ asymmetry then gets converted to 
$B$-asymmetry by the high temperature sphaleron transitions. 
However these sphaleron fields conserve $B-L$. Therefore, the 
produced $B-L$ will not be washed out, rather they will 
keep on changing it to $B$-asymmetry. In a comoving volume 
a net $B$-asymmetry is given by 
\bea
Y_B &=& \frac{n_B}{s}=\frac{28}{79} \epsilon_1 Y_{N1}\delta,
\label{B-asym}
\eea  
where the factor $\frac{28}{79}$ in front~\cite{harvey.90} is the 
fraction of $B-L$ asymmetry that gets converted to $B$-asymmetry. 
Here $\epsilon_{1}$ is given by equation (\ref{cpasym-max1}). 
Further  $Y_{N_1}$ is density of lightest right handed neutrino 
in a comoving volume given by $Y_{N_1}=n_{N_1}/s$, where 
$s=(2\pi^{2}/45)g_{*}T^{3}$ is the entropy density of the Universe 
at any temperature $T$. Finally $\delta$ is the wash out factor at 
a temperature just below the mass scale of $N_1$. The value of 
$Y_{N_1}$ depends on the source of $N_{1}$. For example, 
the value of $Y_{N_1}$ estimated from topological 
defects~\cite{sahuetal.04} can be different from thermal 
scenario~\cite{plumacher.96, buch-bari-plum.02}. In 
the present case we will restrict ourselves to thermal 
scenario only.

Recent observations from WMAP show that the matter-antimatter 
asymmetry in the present Universe measured in terms of 
$(n_B/n_\gamma)$ is~\cite{spergel.03}
\be
\left(\frac{n_{B}}{n_{\gamma}}\right)_{0}\equiv 
\left(6.1^{+0.3}_{-0.2}\right)\times 10^{-10},
\label{baryon-asym}
\ee
where the subscript $0$ presents the asymmetry 
today. Therefore, rewriting equation (\ref{B-asym}) we get 
\be
\left(\frac{n_{B}}{n_{\gamma}}\right)_{0} = 7 (Y_{B})_{0}
  = 2.48 (\epsilon_{1} Y_{N}\delta).
\label{baryon-asym1}
\ee 
Substituting the type-II phase choice relation (\ref{choice-2-max}) 
in (\ref{baryon-asym1}) and comparing with the observed value 
(\ref{baryon-asym}) of the baryon asymmetry we get the bound 
\be
M_1\geq 1.25\times 10^{8} GeV \left( \frac{10^{-2}}{Y_{N_1}\delta}
\right)\left(\frac{0.1eV}{v_L}\right).
\label{M1-bound}
\ee
On the other hand, substitution of $\epsilon_{1,max}$ from the 
type-I phase choice (\ref{cpasym-max1}) in equation 
(\ref{baryon-asym1}) and then comparison with the observed value 
(\ref{baryon-asym}) gives the constraint   
\be
A+B \geq 3.46\times 10^{-3}(10^{-2}/Y_{N}\delta) 
\left(\frac{\left(n_{B}/n_{\gamma}\right)_{0}}
{6.1\times 10^{-10}} \right) \left(\frac{2GeV}{m_\tau}
\right) \left(\frac{v}{174GeV}\right)^{2},
\label{A+B-bound}
\ee
where the physical quantities are normalized with respect 
to their observed values. The above equation, for the values 
of $a,b$ and $c$ from (\ref{abcvalues}), gives only
one constraint on the two hierarchy parameters $\alpha_A$ and
$\alpha_B$. We will determine the individual parameters 
$\alpha_A$ and $\alpha_B$ by demanding that their values 
should reproduce the low energy neutrino parameters correctly,
while satisfying the inequalities $M_1 > O (10^8)$ GeV and 
$\alpha_B > \alpha_A >> 1$.
Individual bounds on $\alpha_A$ and $\alpha_B$ can also be obtained
if we assume that the $\alpha_A$ term and the $\alpha_B$  
term in the sum $A+B$ from equation (\ref{eigenvalue-sum}) are
roughly equal. We then get  
\be
\alpha_{A}=(M_{2}/M_{1}) \leq 17 ~~{\rm and}~~\alpha_{B}=
(M_{3}/M_{1}) \leq 289.
\label{alpha-values-1}
\ee  
\section{Checking the consistency of f-matrix eigenvalues}
%%%%%%%%%%%%%%%%%%%%%%%%%%%%%%%%%%%%
The solar and atmospheric evidences of neutrino
oscillations are nicely accommodated in the minimal
framework of the three-neutrino mixing, in which
the three neutrino flavours $\nu_{e}$, $\nu_{\mu}$, $\nu_{\tau}$
are unitary linear combinations of three neutrino mass eigenstates
$\nu_{1}$, $\nu_{2}$, $\nu_{3}$ with masses $m_{1}$,
$m_{2}$, $m_{3}$ respectively. The mixing among
these three neutrinos determines the structure
of the lepton mixing matrix~\cite{mns-matrix} which
can be parameterized as     
\be
U_{PMNS}=\begin{pmatrix}
c_{1}c_{3} & s_{1}c_{3} & s_{3}e^{i\delta}\\
-s_{1}c_{2}-c_{1}s_{2}s_{3}e^{i\delta} & c_{1}c_{2}-s_{1}s_{2}s_{3}
e^{i\delta} & s_{2}c_{3}\\
s_{1}s_{2}-c_{1}c_{2}s_{3}e^{i\delta} & -c_{1}s_{2}-s_{1}c_{2}s_{3}e^{i\delta} &
c_{2}c_{3}\end{pmatrix} dia(1, e^{i\alpha}, e^{i(\beta +\delta)}),
\label{mns-matrix}
\ee                                                                   
where $c_{j}$ and $s_{j}$ stands for $cos \theta_{j}$
and $sin \theta_{j}$. Here we are interested only in the 
magnitudes of elements of $U_{PMNS}$. Hence for simplicity, we 
neglect all phases in it. The best fit values of the neutrino 
masses and mixings from a global three neutrino flavors oscillation 
analysis are~\cite{gonzalez-garcia_prd.03} 
\be
\theta_{1}=\theta_{\odot}\simeq 34^\circ, ~~\theta_{2}=\theta_{atm}
=45^\circ, ~~\theta_3 \leq 10^\circ,
\label{bestfit-theta}
\ee
and 
\bea
\Delta m_{\odot}^{2}= m_2^2 - m_1^2 & \simeq & m_2^2 = 
7.1\times 10^{-5} \ {\rm eV}^{2}\nonumber\\
\Delta m_{atm}^{2}= m_3^2 - m_1^2 & \simeq & m_3^2 = 
2.6\times 10^{-3} \ {\rm eV}^{2}.
\eea
Using equation (\ref{see-saw}) we rewrite the f-matrix 
\be
f=(\frac{eV}{v_{L}})\left[(m_{\nu}/eV)+\frac{4}{(1.054165)^{2}}
\frac{1}{(M_{1}/{\rm GeV})}
\begin{pmatrix}
\frac{a^{2}}{\alpha_{A}} & 0 & \frac{ab}{\alpha_{A}}\\
0 & a^{2}+\frac{b^{2}}{\alpha_{B}} & \frac{bc}{\alpha_{B}}\\
\frac{ab}{\alpha_{A}} & \frac{bc}{\alpha_{B}} & \frac{b^{2}}
{\alpha_{A}}+\frac{c^{2}}{\alpha_{B}}
\end{pmatrix} \right],
\label{fm1M1}
\ee
where the neutrino mass matrix $m_{\nu}$ is given by 
equation(\ref{diag}). The constrained eigenvalues 
$\alpha_{A}$ and $\alpha_{B}$ are given by equation 
(\ref{alpha-values-1}).

In the following, we choose $M_1$ to be larger than the bound 
given by type-II phase choice (\ref{M1-bound}) and $m_1$ such 
that $m_1^2 << \Delta_{sol}$. For such $m_1$ and $M_1$, we choose 
suitable $\alpha_A$ and $\alpha_B$ that are compatible with the 
low energy neutrino oscillation data. In particular here we 
choose $m_{1}=1.0 \times 10^{-3}eV$, $M_{1}=1.0\times 10^{8}$ GeV, 
$\alpha_{A}=17$, $\alpha_{B}=170$ and $\theta_{3}=6^\circ$. Then 
we get 
\be
f_{dia}=\frac{2.16\times 10^{-3}eV}{v_{L}}
\begin{pmatrix}
1 & 0 & 0\\
0 & 17.3 & 0\\
0 & 0 & 169.7 \end{pmatrix}.
\label{fdia-cal}
\ee 
Thus, for the above values of $m_1$ and $M_1$, the assumed 
hierarchy of right-handed neutrino masses is consistent with 
global low energy neutrino data. Comparing equation (\ref{fdia-cal}) 
with (\ref{fdia-texture}) we get 
\be
\frac{M_{1}}{v_{R}}=\frac{2.16\times 10^{-3}eV}{v_{L}}.
\ee
This implies that $v_{R}=O(10^{10})$ GeV for $v_{L}=0.1$ eV. These 
values of $v_{L}$ and $v_{R}$ are compatible with genuine 
see-saw $v_{L}v_{R}=\gamma v^{2}$ for a small value of 
$\gamma\simeq O(10^{-4})$ \cite{deshpande.91}. 
On the other hand if we choose the parameters $m_{1}=1.0 
\times 10^{-3}$ eV, $M_{1}=1.0 \times 10^{9}$ GeV, $\alpha_{A}=17$, 
$\alpha_{B}=65$ and $\theta_{3}=6^\circ$ we get            
\be
f_{dia}=\frac{1.6 \times 10^{-3}eV}{v_{L}}
\begin{pmatrix}
1 & 0 & 0\\
0 & 16.76 & 0\\
0 & 0 & 64.68 \end{pmatrix}.
\label{fdia-cal-1}
\ee        
Once again we have consistency between the assumed hierarchy 
of right-handed neutrino masses and global low energy neutrino
data. Again comparing equation (\ref{fdia-cal-1}) with 
(\ref{fdia-texture}) we get 
\be
\frac{M_{1}}{v_{R}}=\frac{1.6 \times 10^{-3}eV}{v_{L}}.
\ee
Thus for $v_{L}=0.1$ eV one can get $v_{R}=O(10^{11}$ GeV). Again 
these values are compatible with see-saw for $\gamma\simeq 
O(10^{-3})$. 

Here we demonstrated the consistency of our choice of the 
matrix $f$ with neutrino data for two different choices of 
$\alpha_A$ and $\alpha_B$. For other choices of these 
parameters, to be consistent with $1<<\alpha_A<<\alpha_B$, one 
can choose appropriate values of $m_1\leq 10^{-3}$ eV and 
$M_1\geq 10^8$ GeV in equation (\ref{fm1M1}) which will 
reproduce the correct eigenvalues of the matrix $f$. 

%%%%%%%%%%%%%%%%%%%%%%%%%%%%%%%%%%%%%%%
\section{Conclusion}
In this work we derived an upper bound on the CP-violating 
parameter $\epsilon_{1}$ in left-right symmetric models by 
assuming the case of spontaneous CP-violation. Further we 
assumed a normal mass hierarchy among the heavy Majorana 
neutrinos. A class of left-right symmetric models are 
then considered in which we assume neutrino Dirac masses 
are of the Fritzsch type. We found that keeping the Majorana
phase in type-II mass term of the light neutrinos, gives rise
to a lower bound on the lightest right handed neutrino mass,
whereas keeping the phase in type-I mass term, gives rise to
bounds on the mass ratios $M_2/M_1$ and $M_3/M_1$. We further 
checked that these bounds are consistent with the present 
neutrino oscillation data.

\section{Acknowledgment}
We wish to thank Prof. Urjit A Yajnik for helpful discussions and 
Dr. M C Chen for her useful comments on the manuscript. 

%%%%%%%%%%%%%%%%%%%%%%%%%%%%%%%%%%%%%%%%%%%


\begin{thebibliography}{99}
\bibitem{footnote1}
This conclusion follows immediately from eq. (A.4f) of reference
\cite{mas_npb}, if we substitute $k_2 \ll k_1$.
\end{thebibliography}
\end{document}